\documentclass[referee]{aa} % for a referee version

\usepackage{graphicx}
\usepackage{txfonts}
\def\lsim{\mathrel{\rlap{\lower4pt\hbox{\hskip1pt$\sim$}}
    \raise1pt\hbox{$<$}}}
\def\gsim{\mathrel{\rlap{\lower4pt\hbox{\hskip1pt$\sim$}}
    \raise1pt\hbox{$>$}}}
\begin{document}

  \title{Testing power-law cosmology with galaxy clusters}

%   \subtitle{I. Overviewing the $\kappa$-mechanism}

   \author{Zong-Hong Zhu\inst{1,2}
          \and
	  Ming Hu\inst{1}
	\and
	J.S. Alcaniz\inst{3}
%          \fnmsep\thanks{Just to show the usage} 
	\and
	Yu-Xing Liu\inst{4}
          }

   \offprints{Zong-Hong Zhu}

   \institute{Department of Astronomy, Beijing Normal
	      University, Beijing 100875, China\\
	      \email{zhuzh@bnu.edu.cn}
         \and
	Institut d'Astrophysique Spatiale (IAS),
	CNRS \& Univ. Paris-Sud,
	B\^atiment 121, F-91405 Orsay, France
	\and
	Departamento de Astronomia, Observat\'orio Nacional, 20921-400,
	Rio de Janeiro, Brasil\\
	\email{alcaniz@on.br}
	\and
	The College of Applied Science, Beijing University of Technology,
	Beijing 100022, China
%	\and
%             University of Alexandria, Department of Geography, ...\\
%             \email{c.ptolemy@hipparch.uheaven.space}
%             \thanks{The university of heaven does not accept e-mails}
	}

%   \date{Received September 15, 1996; accepted March 16, 1997}
   \date{Received 00 00, 0000; accepted 00 00, 0000} 

   \abstract
{}
{Power-law cosmologies, in which the cosmological scale factor evolves as a power law in the age, $a \propto t^{\alpha}$ with $\alpha \ga 1$, regardless of the matter content or cosmological epoch, is comfortably concordant with a host of cosmological observations.} {In this article, we use recent measurements of the X-ray gas mass fractions in clusters of galaxies to constrain the $\alpha$ parameter with curvature $k = \pm1, 0$.}
{We find that the best fit happens for an open scenario with the power index $\alpha = 1.14 \pm 0.05$, though the flat and closed model can not be rule out at very high confidence level.} {Our results are in agreement with other recent analyses and show that the X-ray gas mass fraction measurements in clusters of galaxies provide a complementary test to the power law cosmology.}


%   \authorrunning{Zhu, Z. -H, Hu, M., Alcaniz, J.S. \& Liu, Y. -X.}
   \authorrunning{Zhu, Z. -H et al.}

%   \titlerunning{The Power Law Cosmology: Testing with $f_{\rm gas}$}
   \titlerunning{Testing Power Law Cosmology with Galaxy Clusters}

   \maketitle

%    
%________________________________________________________________

\section{Introduction}

Recent measurements of distant supernovae type Ia (SNeIa) suggest that our universe is in an accelerating phase of expansion (Riess et al. 1998; Perlmutter et al. 1999; Astier et al. 2006; Riess et al. 2007). This cosmic acceleration has also been confirmed, independently of the SNeIa magnitude-redshift relation, by the observations of the cosmic microwave background anisotropies (Wilkinson Microwave Anisotropy Probe [WMAP]: Bennett et al. 2003; Durrer et al. 2003; Spergel et al. 2003; 2006) and the large scale structure (LSS) in the distribution of galaxies (Sloan Digital Sky Survey [SDSS]: Tegmark et al. 2006).

The absence of convincing evidence concerning the nature of this phenomenon gave origin to an intense debate and to many theoretical speculations in the last few years. While cosmological constant remains the simplest explanation of cosmic acceleration, it suffers with an impressive fine-tuning and the coincidence problem. (Zeldovich 1968; Weinberg 1989). Given this, dynamical fields are usually invoked, such as quintessence (Ratra \& Peebles, 1988; Frieman et al., 1995; Caldwell et al. 1998; Zhu 1998; Zhu et al. 2001; Sahni \& Starobinsky 2006), phantom fields (Caldwell 2002; Alcaniz 2004; Nesseris \& Perivolaropoulos 2004; Scherrer 2005), quintom (Feng et al. 2005; Wu \& Yu 2005) and Chaplygin gas (Kamenshchik et al. 2001; Bento et al. 2002; Bilic et al. 2002; Dev, Jain \& Alcaniz, 2003; 2004; Zhu 2004; Zhang \& Zhu 2006). There are also some other models where the observed cosmic acceleration is not driven by dark energy. Examples include modified gravity (Perrotta et al. 2000; Riazuelo \& Uzan 2002; Puetzfeld et al. 2005),  theories with extra dimensions (Dvali et al. 2000; Deffayet et al. 2002; Alcaniz 2002; Zhu \& Alcaniz 2005; Alcaniz and Zhu 2005; Pires et al. 2006; Guo et al. 2006), Cardassian cosmology (Freese and Lewis 2002; Zhu and Fujimoto 2002, 2003; Wang et al. 2003), among others (see, e.g., Peebles \& Ratra 2002; Padmanabhan 2003; Copeland et al. 2006; Alcaniz 2006).

In this paper, we explore observational aspects of power-law cosmologies in which the scale factor evolves as $a(t) \propto t^{\alpha}$ with $\alpha \ga 1$. This interesting feature can be obtained when the classical fields are coupled to the curvature of the background space-time in such a way that their contribution to the energy density adjusts itself to cancel the vacuum energy (see, e.g., Ford 1987; Dolgov 1997). Such a scaling relation can also come from a SU(2) cosmological instanton-dominated universe (Allen 1999; Dev et al. 2001; Sethi et al. 2005). The power-law cosmology model is now becoming particularly interesting since it is a generic feature in a class of models that attempt to dynamically solve the $\Lambda$ problem. From the observational viewpoint, it has been shown that the age of the linear coasting universe ($\alpha = 1$) is $\sim 50\%$ larger than the age of standard CDM universe (Lohiya \& Sethi 1999),  which makes it comfortably concordant with the ages of globular clusters.
It can also be demonstrated that this model is consistent with the primordial nucleosynthesis  (Batra et al. 1999).

Power-law cosmologies have triggered a wave of interests aiming to constrain its model parameter using various cosmological observations, such as the magnitude-redshift relation of supernovae of type Ia (Dev et al. 2001; Sethi et al. 2005), gravitational lensing statistics (Dev et al. 2002), the angular size - redshift data of compact radio sources (Jain, Dev and Alcaniz 2003), the present age of the universe (Kaplinghat et al. 1999), the age measurements of high-$z$ objects (Dev et al. 2002; Sethi, Dev and Jain 2005), and the primordial nucleosynthesis (Kaplinghat et al. 1999; Batra, Sethi and Lohiya 1999; Kaplinghat, Steigman and Walker 2000). In this work, we shall consider the observational constraints on the model parameter of the power law cosmology arising from X-ray gas mass fractions measurements of clusters of galaxies, as recently discussed by Allen et al. (2002, 2003, 2004). We find that the best-fit happens for the open model with the power index $\alpha = 1.14 \pm 0.05$, though the flat and closed model can not be   rule out at very high confidence level. Our results are in agreement with other recent analyses, providing a complementary test to this general class of models. The plan of the paper is as follows. In the next section, we provide a brief summary of power-law cosmologies and basic equations relevant to our work. Observational constraints from X-ray gas mass fractions of clusters of galaxies are discussed in Section III. we end this paper by summarizing our main results and conclusions in Section IV.

%__________________________________________________________________

\section{Basics of power-law cosmologies}

Let us first consider a homogeneous and isotropic universe described by the Friedman-Robertson-Walker (FRW) line element,
\begin{equation}
ds^2=dt^2-a^2(t)\left[\frac{dr^2}{1-k
r^2}+r^2(d\theta^2+\sin^2\theta d\phi^2)\right]\;,
\end{equation}
Where $t$ is cosmic proper time, $a(t)$ is the scale factor and $r, \theta, \phi$ are commoving spherical coordinates. have relation $a(t) = a_0[1 + z]^{-1}$ where $z$ is the redshift, and  $a_0$ is the current value of the scale factor.

We study a general power-law cosmology with the scale factor given by 
\begin{equation}
a(t) = \kappa t^{\alpha}\;,
\end{equation}
where $\kappa$ and $\alpha$ are two parameters.

The expansion rate of the universe is described by a Hubble parameter, \begin{math}H(t)=\dot{a}/a = \alpha/t\end{math}, so that the present expansion rate of the universe is \begin{math}H_0=\alpha/t_0\end{math} (here and subsequently the subscript 0 on a parameter refers to its present value).
We set $h=H_0/100$ km$\cdot$s$^{-1}\cdot$Mpc$^{-1}$, i.e., the Hubble constant in units of 100 km$\cdot$s$^{-1}\cdot$Mpc$^{-1}$.
The scale factor and the redshift are related to their present values by \begin{math}a/a_0=(t/t_0)^\alpha\end{math}.

From the above Eqs., we can easily express the angular diameter distance as:
\begin{equation}
D^A=\left(\frac{\alpha}{H_0}\right)^{\alpha}(1+z)^{-1}\times
{\rm sinn}\left[\frac{1}{1-\alpha}\left(\frac{\alpha}{H_0}\right)^{1-\alpha}(1-(1+z)^{1-\frac{1}{\alpha}}\right],
\end{equation}
where for closed model, \begin{math}k=1\end{math}, \begin{math}{\rm sinn}(x)=\sin(x)\end{math}; for flat model, \begin{math}k=0\end{math}, \begin{math}{\rm sinn}(x)=x\end{math}; and for open model, \begin{math}k=-1\end{math}, \begin{math}{\rm sinn}(x)=\sinh(x)\end{math}.
In the case of a linear coasting universe, i.e., $\alpha = 1$, Eq.(2) takes the following form,
\begin{equation}
D^A=\frac{1}{H_0(1+z)}\times {\rm sinn}\left[\ln(1+z)\right]\;,
\end{equation}
where the luminosity-distance $D^L$ is related to the angular diameter as \begin{math}D^A=D^L(1+z)^{-2}\end{math}.

\begin{figure}
   \centering
   \includegraphics[width=8.9cm]{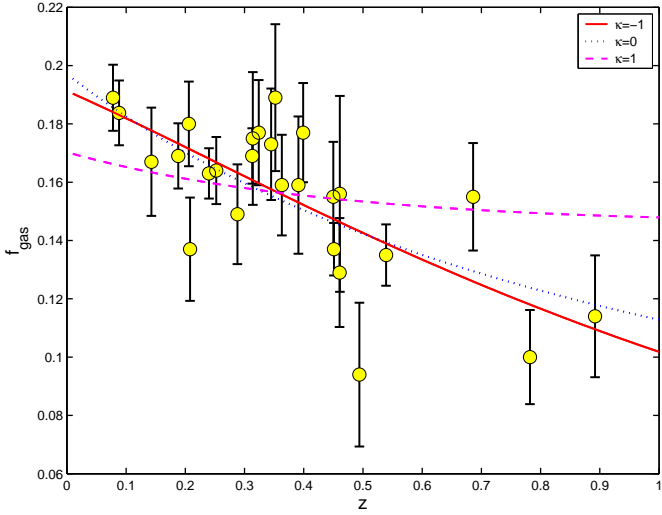}
   \caption{The measurements of $f_{\rm gas}$ for 26 rich clusters using
  standard cold dark matter cosmology.
  The red curve corresponds to our best fit to the open model
  ($k=-1$) with $\alpha=1.14$.
  The blue curve corresponds to our best fit to the flat model $k=0$ with
  $\alpha=2.3$, and the magenta curve corresponds to our
  best fit to the closed model ($k=+1$) with $\alpha=0.95$.}
\label{fig:h2}
    \end{figure}
%__________________________________________________________________

\section{Constraints from the X-ray gas mass fraction of galaxy clusters }

By considering the usual assumption that the rich clusters are large enough to provide a fair sample of the matter content of the whole universe (White et. al. 1993), the ratio of the baryonic content to total mass in clusters, $f_{\rm gas}$, should be the same as the ratio of the cosmological parameters, $\Omega_b/\Omega_m$, where $\Omega_b$ and $\Omega_m$ are the mean baryonic and total mass densities of the universe in units of the critical density (Allen et al. 2002). The measured $f_{\rm gas}$ and $\Omega_b$ allow us, therefore, to determine $\Omega_m$.
Sasaki(1996) showed how the measurements of $f_{\rm gas}$ in rich clusters 
%(and not on those that might be in the process of collapse) at
at different redshifts can provide an efficient way to constrain cosmological parameters decribing the geometry of the universe.
This is based on the fact that the measured $f_{\rm gas}$ values for each rich cluster depend on the angular diameter distances as $f_{\rm gas} \propto [D^A]^{3/2}$. The underlying cosmology should be the one which places the rich clusters at the right angular diameter distances to ensure measured $f_{\rm gas}$ values to be invariant with redshift (Sasaki 1996; Allen et al. 2002).

Using the {\it Chandra} observational data, Allen et al. (2002, 2003, 2004) have obtained the $f_{\rm gas}(r)$ profiles (the gas mass fraction value within the radius of $r$) for the 26 relaxed clusters, that are increased with $r$ in the inner parts of clusters. However, except for Abell 963, the $f_{\rm gas}(r)$ profiles of the other clusters appear to have converged or be close to converging with a radius $r_{2500}$, which is defined as the radius within which the mean mass density is 2500 times the critical density of the universe at the redshift of the cluster (Allen et al. 2002, 2003, 2004). The gas mass fraction values of these 26 clusters at $r_{2500}$ (or at the outermost radii studied for PKS0745-191 and Abell 478) within the framework of standard cold dark matter (SCDM) cosmology are shown in Figure 1. We will use this database to constrain the power law cosmological model.
Following Allen et al. (2002) (see also Lima et al. 2003), we have the model function as
\begin{equation}
f_{\rm gas}^{\rm mod}(z_i;\alpha) =
      \frac{ b \Omega_b}{\left(1+0.19{h}^{1/2}\right) \Omega_m}
  \left[{h\over 0.5}
        \frac{D^A_{\rm{SCDM}}(z_i)}{D^A_{\rm lin}(z_i;\alpha)}
                \right]^{3/2}
\end{equation}
where the bias factor $b$ is a parameter motivated by gas dynamical simulations, which suggest that the baryon fraction in clusters is slightly depressed with respect to the Universe as a whole (Eke et al. 1998; Bialek et al. 2001). The term $(h/0.5)^{3/2}$ represents the change in the Hubble parameter from the defaut value of $H_0 = 50 {\rm{km \, s^{-1} \, Mpc^{-1}}}$ and the ratio ${D^A_{\rm{SCDM}}(z_i)}/{D^A_{\rm{lin}}(z_i;\alpha)}$ accounts for the deviations of the power law cosmology from the default SCDM cosmology.

   \begin{figure*}
   \centering
   \includegraphics[width=6.0cm]{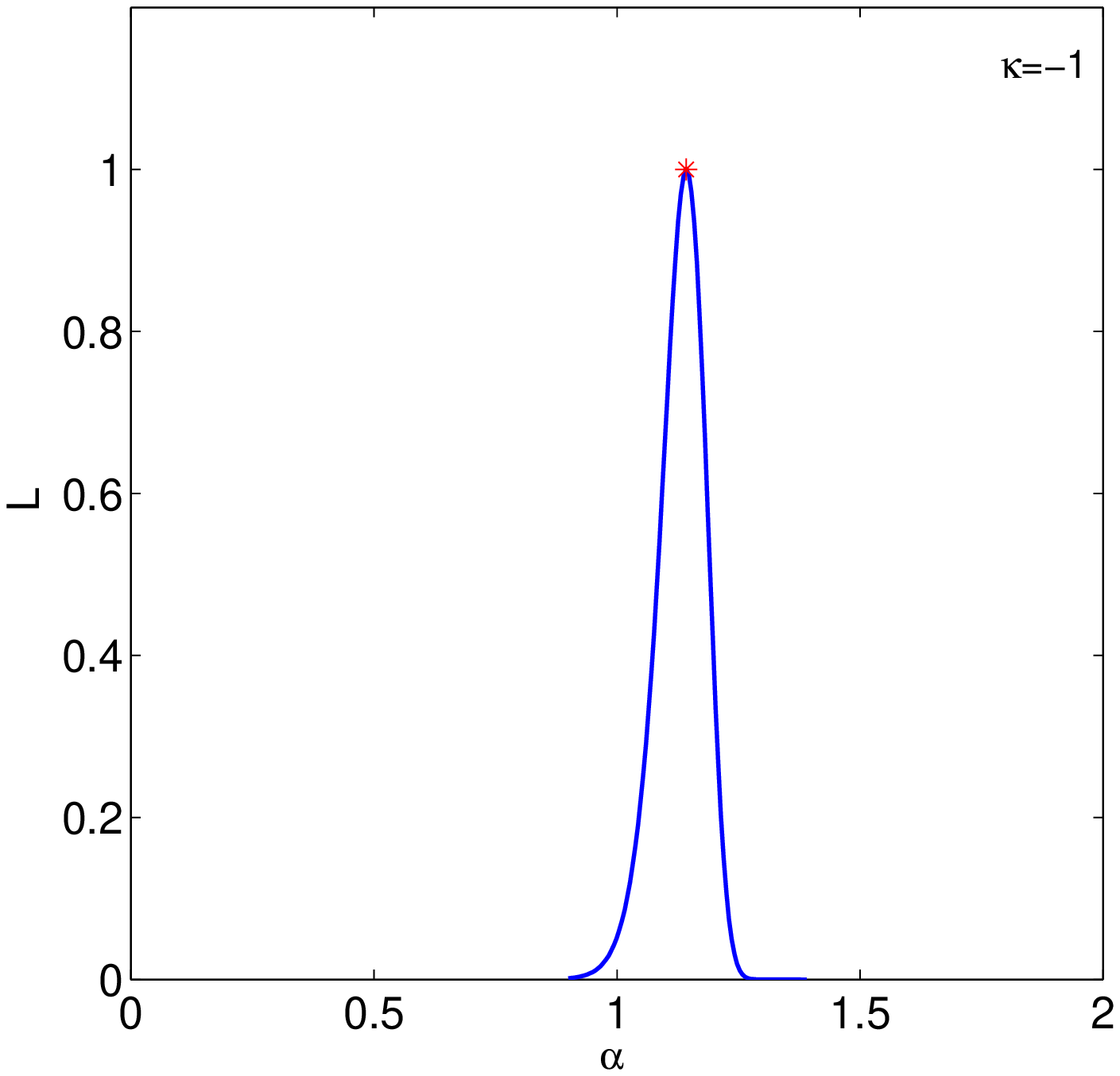}
   \includegraphics[width=6.0cm]{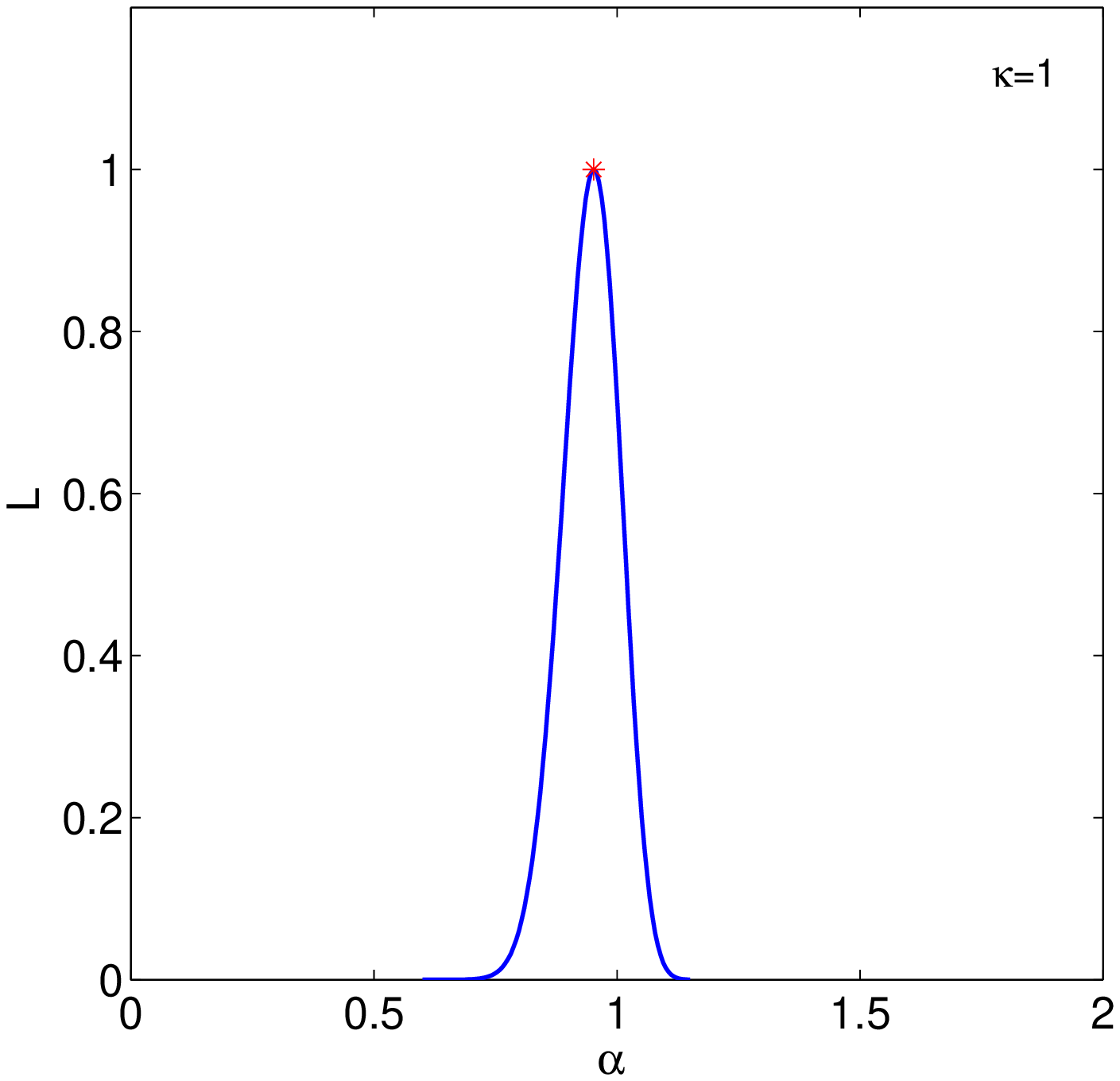}%\\
%   \begin{center}
   \includegraphics[width=6.0cm]{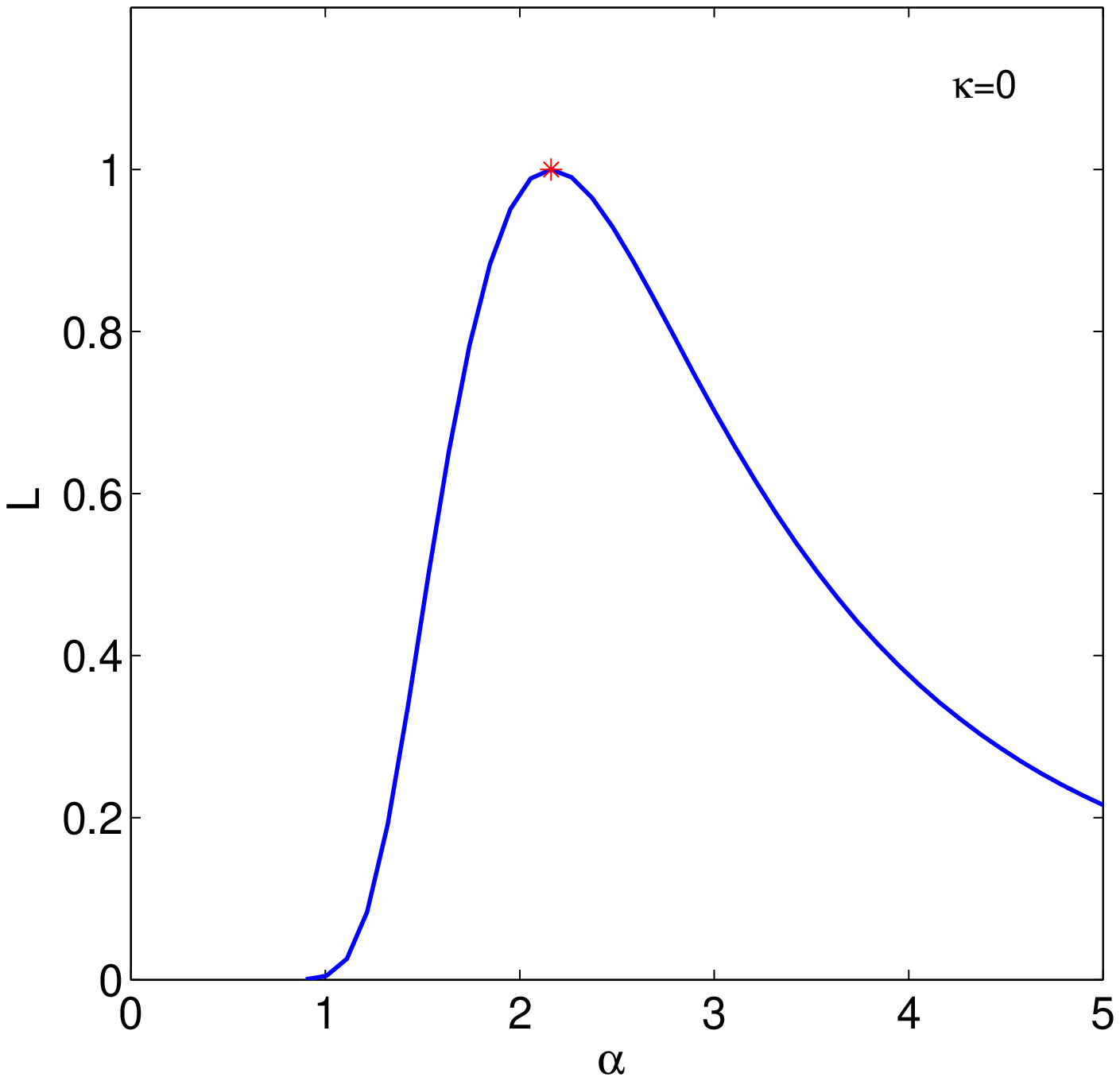}
%   \end{center}
\caption{Marginalized likelihood functions (normalized at 1 at
  maximum) for the open ($k=-1$), closed ($k=+1$) and flat ($k=0$) models
  using X-ray gas mass fraction data of 26 rich galaxy clusters.}
\end{figure*}

\subsection{Analysis}

We now use the \begin{math}f_{\rm gas}\end{math} values of the 26 clusters to constrain the parameter $\alpha$ of the power law cosmology model through a $\chi^{2}$ minimization method. Following Allen et al. (2002, 2003), uncertainties of \begin{math}\Omega_bh^2\end{math}, $b$ and $h$ are accounted for by using Gaussian priors, as \begin{math}\Omega_bh^2=0.0214\pm0.002\end{math}, \begin{math}b=0.824\pm0.089\end{math} and $h = 0.72 \pm 0.08$. Galaxy dynamics presents constraints on $\Omega_m$ independent of cosmological models. Here we invoke the best-fit value of $\Omega_m$ of a million galaxy of SDSS (Blake et al. 2006), $\Omega_mh=0.20\pm 0.03$. The $\chi^2$ difference between the model function and the data determined by using SCDM cosmology is then
%
%\begin{equation}
%\chi^2(\alpha) = \sum_i\frac{[f_{\rm gas}^{\rm mod}(\alpha,z_i)-f_{{\rm gas},\
%i}]^2}{\sigma_{f_{\rm gas},\
%i}^2}+\left[\frac{\Omega_bh^2-0.0214}{0.002}\right]^2+\left[\frac{b-0.824}{0.089}\right]^2 +
%  \left[\frac{h - 0.72}{0.08}\right]^{2} +
%  \left[\frac{\Omega_mh-0.20}{0.03}\right]^2
%\end{equation}
%%%%%%%%%%%%%%%%%%%%%%
\begin{eqnarray}
\chi^2(\alpha) = & \sum_i\frac{[f_{\rm gas}^{\rm mod}(\alpha,z_i)-f_{{\rm gas},\
i}]^2}{\sigma_{f_{\rm gas},\
i}^2}+\left[\frac{\Omega_bh^2-0.0214}{0.002}\right]^2   \nonumber\\
& +\left[\frac{b-0.824}{0.089}\right]^2 +
  \left[\frac{h - 0.72}{0.08}\right]^{2} +
  \left[\frac{\Omega_mh-0.20}{0.03}\right]^2
\end{eqnarray}
where $f_{\rm gas}^{\rm mod}(z_i;\alpha)$ refers to equation~(4), $f_{{\rm gas,o}i}$ is the measured $f_{\rm gas}$ with the default cosmology (SCDM and $H_0 = 50 {\rm{km \, s^{-1} \, Mpc^{-1}}}$) and $\sigma_{f_{{\rm gas},i}}$ is the symmetric root-mean-square errors ($i$ refers to the $i$th data point, with totally 26 data). The summation is over all of the observational data points. The probability distribution of \begin{math}\alpha\end{math} are determined by marginalizing the nuisance parameters, $L(\alpha)=\int\exp(-\chi^2/2)\ {\rm d}b\ {\rm d}h\ {\rm d}(\Omega_bh^2)\ {\rm d}(\Omega_mh)$.
The integral is over all the range of $b$, $h$, $\Omega_bh^2$ and $\Omega_mh$.

The results of our analysis using X-ray gas data are displayed in Figure 2. In this figure we show marginalized likelihood functions (normalized at 1 at maximum) for open, closed and flat models respectively. The best fit values for $\alpha$ corresponds to the maximum $L$. A horizontal line of any value of $L$ will intersect two values, $\alpha_1$ and $\alpha_2$, which define an integral range of $L$. If this integral is 68.3\% of the area that the $L$ curve embodies, the $\alpha_1$ and $\alpha_2$ will be the lower and upper bounds of the power index at the confidence of $1\sigma$, respectively. The numerical results are shown in Table 1.

\begin{table}[h]
\begin{center}
\begin{tabular*}{8cm}{c|c|c}
\hline
$k$ & $\alpha$ & $\chi^2_{min}$ \\
\hline
 \ \ \ \ \ -1\ \ \ \ \  &\ \ \ \ \  \begin{math}1.14\pm0.05\end{math}\ \ \ \ \ &\ \ \ \ \  22.9\ \ \ \ \ \\
\hline
 0 & \begin{math}2.3_{-0.7}^{+1.4}\end{math} & 23.9\\
\hline
 1 & \begin{math}0.95\pm0.06\end{math} & 45.0\\
\hline
\end{tabular*}
\end{center}
\caption{Best-fit values for power-law cosmologies using X-ray gas mass of clusters.}
\end{table}

%__________________________________________________________________

\section{Final Remarks}

%The increasing bulk of astrophysical data accumulated in recent years
%  has delineated a new standard cosmological paradigm.
%According to this picture, a linear coasting cosmology has been proposed.
%It can solve the "$\Lambda$ problem", the "horizon problem" and
%  the "age problem".
%Comparing with other cosmology models such as cosmological constant
%  \begin{math}\Lambda\end{math} model or other scalar field
%  quintessence, linear coasting cosmology is completely different in
%  the dynamical properties, but as the other models, they all share
%  the ability of well fitting the same data set.
The increasing bulk of astrophysical data accumulated in recent years has suggested that we live in a flat, accelerating universe. This cosmic acceleration has been attributed to a dark energy component with negative pressure. While the simplest model for dark energy is the cosmological constant, it suffers from a fine-tuning and coincidence problems (Zeldovich 1968; Weinberg 1989). In this paper, we explored power-law cosmologies which have the potential of explaining various cosmological data, such as supernovae of type Ia, the age of the universe and the primordial nucleosynthesis.. 

In this paper, we have used the X-ray mass fraction data of galaxy clusters, $f_{gas}$, to constrain the parameters of the power law cosmology.
The results are shown in table 1 and Figure 2. From this analysis, we find that the best-fit model happens for the open model with the power index
$\alpha = 1.14 \pm 0.05$. However, the flat model has almost the same value of the minimum $\chi^2$ as
the open model, though the error of $\alpha$ is slight larger.
This is in accordance with the result of Dev et al. (2001). Our analysis, therefore, provides an independent and complementary test to this class of model.

Finally, we emphasize that power-law cosmologies are an interesting alternative for describing the universe and seem to be able to solve theoretical and observational matters (as, e.g., the age and horizon problem) if the power index $\alpha \ga 1$. We believe that future data should provide a much tighter constraint to this class of cosmological scenario.

%__________________________________________________________________
\begin{acknowledgements}
We would like to thank 
  N. Aghanim for helpful discussion 
	and
  S. Allen for sending us their compilation of the X-ray mass fraction data.
%%%
%{\bf
Our thanks go to the anonymouse referee for valuable comments and useful
  suggestions, which improved this work very much.
%}
%%%
This work was supported by
  the National Natural Science Foundation of China, under Grant No. 10533010,
  973 Program No. 2007CB815401, Program for New Century Excellent Talents in
  University (NCET) of China
  and the Project-sponsored by SRF for ROCS, SEM of China.
Z.-H. Z. also acknowledges support from CNRS,
  and is grateful to all members of cosmology group at IAS for their
  hospitality and help during his stay.
J.S. Alcaniz is supported by FAPERJ under Grant No. E-26/171.251/2004 and
  CNPq  under Grant No. 307860/2004-3;  475835/2004-2;  485662/2006-0.
\end{acknowledgements}

\end{document}